\documentclass[11pt,a4paper]{article}
\usepackage{amsmath,amssymb}
\usepackage{epsfig,graphicx}

\begin{document}

\title{\bf Neutrino magnetic moment in a magnetized plasma}
\author{N.~V.~Mikheev$^a$\footnote{{\bf mikheev@uniyar.ac.ru} },
E.~N.~Narynskaya$^{a}$\footnote{{\bf elenan@uniyar.ac.ru}}
\\
$^a$ \small{\em Yaroslavl State University} \\
\small{\em  150000 Russian Federation, Yaroslavl, Sovietskaya 14}
}
\date{}
\maketitle

\begin{abstract}
The contribution of a magnetized plasma to the neutrino 
magnetic moment is calculated. It is shown that only part of the
 additional neutrino  energy
 in magnetized plasma connecting with its spin and magnetic field strength 
defines the neutrino  magnetic moment. It is found that 
the presence of magnetized plasma  does not lead to the considerable increase of the neutrino
magnetic moment in contrast to the 
results presented  in literature previously.
\end{abstract}

\section{Introduction}
\indent\indent
Investigations of the influence of an active medium 
on neutrino dispersion are based on calculations of the 
neutrino self-energy operator $\Sigma(p)$, which is defined in the terms of the  invariant amplitude of the neutrino forward scattering by the relation
\begin{equation}
M_{(\nu \to \nu)} = - \, {\bar U(p)} \, \Sigma(p)\,U(p). \label{eq:1}
\end{equation}

It should be noted that the studies of this operator in external active medium, which  can be presented by the magnetic field as well as dense plasma, have a long history.  So,  the  calculations of the neutrino self-energy operator in a magnetized plasma were presented in \cite{Olivo}-\cite{Elizalde}.

The great  interest to the neutrino self-energy operator in an active medium is caused  by the possibility  of the production from it the  data on neutrino magnetic moment. In particular, such investigation were performed in the paper \cite{Zhykovski} under physical conditions
\begin{equation}
E_\nu \ll \frac{m^2_W}{\mu}, \label{eq:2}
\end{equation}
where $E_\nu$ is the neutrino energy, $m_W$ is the mass of W-bozon, $\mu$ is the chemical potential of a plasma. 

The additional neutrino magnetic moment  was calculated  in the paper \cite{Zhykovski}  in the limit of strong magnetic field, 
$2eB > \mu^2 - m^2_e$,
\begin{equation}
\Delta \mu_\nu = \frac{16 \pi^3}{3}\, \frac{n_e \, (E_\nu - q_3)}{m^2_W \,  m^2_\nu} \, \mu^{(0)}_\nu,  \label{eq:3}
\end{equation}
and in the weak magnetic field limit, $2eB \ll \mu^2 - m^2_e$,
\begin{equation}
\Delta \mu_\nu = -\frac{8}{3}\, \frac{(3\, \pi^2 \, n_e)^{1/3} }{( E_\nu + q_3)\,m_\nu} \, \mu^{(0)}_\nu,\label{eq:4}
\end{equation}

where $\mu^{(0)}$ is the neutrino magnetic moment in vacuum 
\cite{Lee, Shrok}: 
$$
\mu^{(0)}_\nu = \frac{3e G_F}{8 \sqrt 2 \, \pi^2} \,m_\nu.
$$

As one can see the expression (\ref{eq:3})  consists the enhancement by the neutrino mass $m_\nu$ in denominator. While the second result in the limit of weak magnetic field is independent on its mass. Such result as it was pointed in paper \cite{Elmfors} are somewhat misleading, so it is interest to perform new independent investigation of neutrino magnetic moment in a magnetized plasma.

 As noted above  for such investigation we need in neutrino self-energy operator $\Sigma(p)$. In a magnetized plasma this operator can be presented in the general form:
   \begin{eqnarray}
\Sigma(p) & = & [A_L \, (p \,\gamma) + B_L \, (u \, \gamma) + C_L \, (p \tilde F \gamma) ]\, \gamma_L + 
\nonumber
\\[2mm]
& + & [A_R \, (p \, \gamma) + B_R \, (u \, \gamma) + C_R \, (p \tilde F \gamma) ]\, \gamma_R +
m_\nu [K_1 + iK_2 (\gamma F \gamma)], \label{eq:5}
\end{eqnarray}
where $u^\mu$ is the 4-vector of medium velocity, $p^\mu$ is the neutrino 4-momentum, $\gamma_{L,R} = (1 \mp \gamma_5)/2$, $A_R, B_R, C_R, A_L, B_L, C_L, K_1, K_2$ are the numerical coefficients, which in general case can be depend on the magnetic field strenght and spin neutrino, $F^{\mu\nu}$ and $\tilde F^{\mu\nu}$ are the tensor and dual tensor of the electromagnetic field. 

A change of the neutrino energy caused by its forward scattering in a medium can be expressed via the 
neutrino self-energy operator as follows
\begin{equation}
\Delta E_\nu = \frac{1}{4E_\nu}\, Sp \left \{ ((p\gamma) + m_\nu)\,(1 -  (s\gamma)\,\gamma_5)\,\Sigma (p) \right \}. \label{eq:6}
\end{equation}

Taking into account the general expression for neutrino self-energy operator $(\ref{eq:5})$  one can  rewrite $\Delta E_\nu$ in the form 

 \begin{eqnarray}
 \Delta E_\nu & = & \frac{m^2_\nu}{2 E_\nu} \, (A_L + A_R +2K_1) + B_L \, \frac{1 - (\vec \xi \cdot \vec  v)}{2} + B_R \, \frac{1 + (\vec \xi \cdot \vec  v)}{2} -
\nonumber \\[2mm]
 & - &    {\frac{m_\nu}{2} \, (C_L - C_R + 4K_2) \,
 [(\vec \xi \cdot \vec B_t) + \frac{m_\nu}{E} \, (\vec \xi \cdot \vec B_l)]}. \label{eq:7}
 \end{eqnarray}
Here $E_\nu$ is the neutrino energy in a vacuum, 
$\vec \xi$ is the double average neutrino spin vector,
${\vec  B_l}$ and ${\vec B_t}$ are the longitudinal and transverse magnetic field components relative to the direction of neutrino propagation respectively, 
${\vec v}$ is the neutrino velocity vector.

In order to obtain the additional neutrino energy due to presence of the neutrino magnetic moment only, one has to use the lagrangian

\begin{equation}
\Delta L^{(\mu)}_{int} = \frac{i\mu_\nu}{2} \, (\, \bar \Psi \, \sigma_{\mu\nu} \, \Psi \,) \, F^{\mu\nu}, \label{eq:8}
\end{equation}
where $\Psi$ is the fermion field, $\sigma_{\mu\nu} = (\gamma_\mu\,\gamma_\nu - \gamma_\nu \, \gamma_\mu)/2$.

Substituting this lagrangian into the expression for the additional neutrino energy

\begin{equation}
\Delta E_\nu^{(\mu)} =  - \int dV \, < \, \Delta L^{(\mu)}_{int} \, >, 
\label{eq:9}
\end{equation}
one can obtained the contribution to neutrino energy caused by the neutrino magnetic moment in the form
\begin{equation}
\Delta E_\nu^{(\mu)} = - \mu_\nu  \,[(\vec \xi \cdot \vec B_t) + \frac{m_\nu}{E} \, (\vec \xi \cdot \vec B_l)].\label{eq:10}
 \end{equation}

 A comparison of expression (\ref{eq:7}) with the result (\ref{eq:10}) shows that only last term in the expression for total additional neutrino energy (\ref{eq:7}) corresponds to the  neutrino magnetic moment. So the neutrino magnetic moment can be written as follows:
 \begin{equation}
 \mu_\nu = \frac{m_\nu}{2} \, (C_L - C_R + 4K_2) \label{eq:11}
 \end{equation}
 and defined by the coefficients $C_R$, $C_L$ and $K_2$.

 The purpose  of our paper is the calculation of the neutrino magnetic moment induced by the magnetized plasma.

\section{Magnetic moment of neutrino in a magnetized plasma}
\indent\indent
In this section, we calculate the terms of the neutrino self-energy operator $\Sigma(p)$, which give contribution to 
the magnetic moment of neutrino. 

In the magnetized plasma neutrino magnetic moment can be presented as a sum of field and plasma contributions
\begin{equation}
\mu_\nu = \mu_\nu^{field} + \mu_\nu^{plasma}. \label{eq:12}
\end{equation}

The field contribution to the neutrino magnetic moment is widely discussed in literature. The expression for $\mu_\nu^{field}$ in a broad range of neutrino energy and magnetic field under conditions
$$ m^2_l/m^2_W \ll (eB)^2p^2_\perp/m^6_W \ll 1$$
 can be extracted, for example, from the paper \cite{Mikheev}
  \begin{equation}
  \mu_{\nu_l}  \simeq  \mu^0_{\nu_l} \left \{ 1 + \frac{4 \, \chi^2}{3} \left ( \ln \frac{1}{\chi} - \frac{17}{3} + \ln 3 + 2 \, \gamma_E + i\pi \right )\right \}. \label{eq:13}
  \end{equation}
Here $p^2_\perp = \sqrt{p_x^2 + p^2_y}$, $\chi^2 = (eB)^2p^2_\perp/m^6_W $, $\lambda = m^2_l/m^2_W$, $\gamma_E = 0,577\cdots$  is the Euler constant.

In order to find the plasma contribution to the neutrino magnetic moment it is necessary to calculate the coefficients $C_L$, $C_R$ and $K_2$ of the neutrino self-energy operator. 

We consider  real astrophysical conditions, when the mass of the mass of $W$-boson is the largest physical parameter
\begin{equation}
 m^2_e, \mu^2, T^2, eB \ll m^2_W, \label{eq:cond}
 \end{equation}
where $\mu$ and $T$ are the chemical potential and temperature of the plasma respectively.

 Under these conditions the main contribution to the amplitude of the neutrino forward scattering (and, hence, the self-energy operator) is caused by the neutrino scattering on  plasma electrons and positrons. The amplitude of the 
 neutrino forward scattering in a magnetized plasma can be 
presented as the sum of three terms that correspond 
to three diagrams depicted in the fig.1:
\begin{figure}[t]
\centerline{\includegraphics{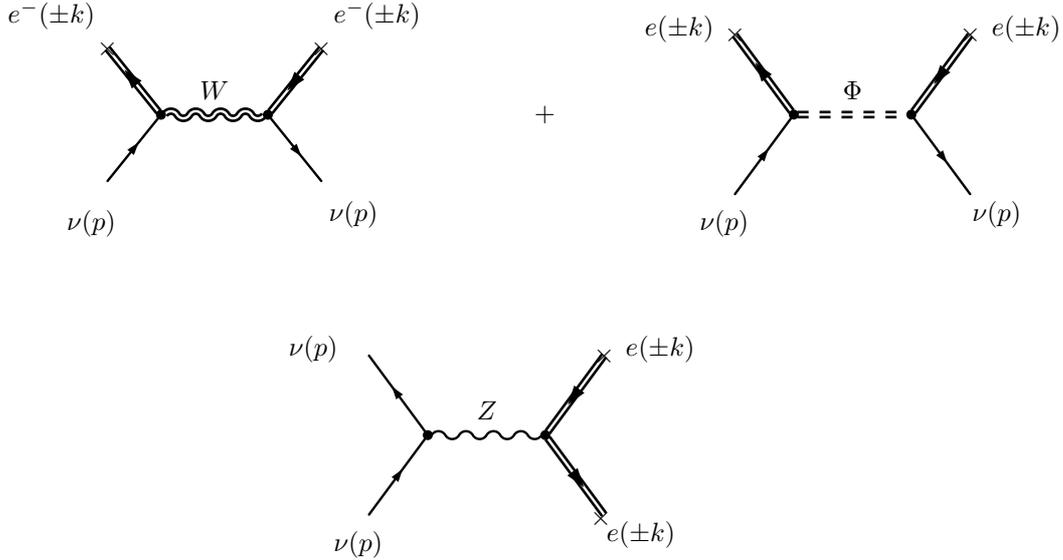}} 
\caption{Feynman diagrams determining the contribution of a magnetized plasma to the amplitude of neutrino forward scattering. Double lines correspond to charged particles.}
\label{fig:fig1}
\end{figure}
$$ \Sigma(p) = \Sigma^W(p) + \Sigma^Z(p) + \Sigma^\Phi(p). $$

The process of neutrino scattering  $\nu \to \nu$ via exchange of $W$-bozon is caused by the lagrangian
 \begin{equation}
 L = \frac{g}{2\sqrt{2}} \left ( \bar \Psi_e \,\gamma_\alpha\,(1 - \gamma_5)\,
\Psi_{\nu_e} \, \right )\,W_\alpha + \frac{g}{2\sqrt{2}}
 \left ( \bar \Psi_{\nu_e} \,\gamma_\alpha\,(1 - \gamma_5)\,
\Psi_{e} \, \right )\,W^*_\alpha.      
       \label{eq:lag-1}
       \end{equation}

 Here $\Psi_e$, $\Psi_{\nu_e}$, $W_\alpha$ are the quantized field of electron, neutrino and $W$-bozon correspondingly,  $g$ is the electroweak coupling constant in Standard Model. 

The amplitude of this process and its contribution to the neutrino self-energy operator 
 $\Sigma(p)$ was studied in detail previously \cite{Mikheev:2010}:
\begin{eqnarray}
\Sigma^W(p) & = & \frac{i\,g^2}{2} \, \sum^{\infty}_{n=0} (-1)^n \, \int \frac{d^3k}{(2\pi)^3} \, \frac{e^{-z}}{\omega_n} \times
\label{eq:16}\\[2mm]
& \times & \left (   f(\omega_n) \, G^W_{\beta \alpha} (p-k) - \tilde f(\omega_n) \, G^W_{\beta \alpha} (p+k) \right) \times \nonumber \\[2mm]
& \times & \gamma_\alpha \, 
[\, \hat k_\parallel (L_n(2z)\, \Pi_{-} - L_{n-1}(2z)\,\Pi_{+}) + 2 \hat k_\perp \, L^1_{n-1} (2z)\, ] \, \gamma_\beta \, \gamma_L , \nonumber
\end{eqnarray}
where $z = k^2_\perp/\beta$, $\beta = eB$,  $\hat k_\parallel = (k \tilde \varphi \tilde \varphi \gamma) = \omega_n \gamma_0 - k_3\gamma_3$, $ \hat k_\perp = (k \varphi \varphi \gamma)=k_1\gamma_1 + k_2\gamma_2$,
  $\varphi_{\alpha\beta}= F_{\alpha \beta}/B$ and   $\tilde \varphi_{\alpha\beta} =  
  \varepsilon_{\alpha\beta\rho\sigma} \varphi_{\rho\sigma}/2$  are 
dimensionless tensor and dual tensor of external magnetic field~\footnote{The magnetic field is directed along the z-axis $\vec B = (0, 0, B)$.},
$\Pi_{\pm} = \frac{1}{2}(1 \pm  i \gamma_1 \gamma_2)$ are the projection operators,
  $L_{k}^{l}(x)$ are the associated Laguerre polynomials,
$G^W_{\beta \alpha}(q)$ is the Fourier image of the translation-invariant part of the W-boson propogator, $f(\omega_n)$  and $\tilde f(\omega_n)$ are the distribution function of electrons and positrons respectively. In the plasma rest frame these functions have the following form
 $$ f(\omega_n) = [e^{(\omega_n - \mu)/T} + 1]^{-1}, \,\,\, \tilde f(\omega_n) = [e^{(\omega_n + \mu)/T} + 1]^{-1},$$
where 
$\omega_n$ is the energy of electron (positron) on the Landau level with number $n$.

For the contribution from the process of neutrino scattering
 due to the exchange of charges scalar $\Phi$-boson, one has to use the lagrangian
 \begin{eqnarray}
 L & = & -\frac{g}{2\sqrt{2}}  \bar \Psi_e \,
\left ( \frac{m_e - m_{\nu}}{m_W} - \frac{m_e +  m_{\nu}}{m_W} \, \gamma_5 \,
\right )\,\Psi_{\nu_e} \, \Phi - \nonumber \\
& \qquad & \qquad - \frac{g}{2\sqrt{2}}  \bar \Psi_e \,
\left ( \frac{m_e - m_{\nu}}{m_W} - \frac{m_e + m_{\nu}}{m_W} \, \gamma_5 \,
\right )\,\Psi_{\nu_e} \, \Phi^*. 
       \label{eq:lag-2}
       \end{eqnarray}
and have the following result:
\begin{eqnarray}
\Sigma^\Phi(p) & = & - \frac{i\,g^2}{2} \, \sum^{\infty}_{n=0} (-1)^n \, \int \frac{d^3k}{(2\pi)^3} \, \frac{e^{-z}}{\omega_n} \times
\label{eq:17}\\[2mm]
& \times & \left (   f(\omega_n) \, G^\Phi (p-k) - \tilde f(\omega_n) \, G^\Phi (p+k) \right) \times \nonumber \\[2mm]
& \times & \left \{  \frac{m^2_e \, m_\nu}{m^2_W} \, (L_n(2z)\,\Pi_{-} - L_{n-1}(2z)\,\Pi_{+}) - \right . \nonumber \\[2mm]
& - &  [\hat k_\parallel (L_n(2z)\, \Pi_{-} - L_{n-1}(2z)\,\Pi_{+}) +  \nonumber \\[2mm]
& + & \left . 2 \hat k_\perp \, L^1_{n-1} (2z)\, ] \,
\left (\frac{m^2_e}{m^2_W} \, \gamma_L - 
\frac{m^2_\nu}{m^2_W}\,\gamma_R \right )\, \right \}.
 \nonumber
\end{eqnarray}

Here $G^\Phi_{\beta \alpha}(q)$ is the Fourier image of the translation-invariant part of the $\Phi$-boson propagator.

Note, that the contributions 
(\ref{eq:16}) and (\ref{eq:17}) correspond to the electron neutrino only, for the other type of neutrino, $\nu_\mu, \nu_\tau$, the part of neutrino self-energy operator caused by the process with  the exchange of charges scalar $\Phi$-boson is zero,  $\Sigma^W(p) = \Sigma^\Phi(p) =0$.

Under physical conditions considered (\ref{eq:cond}) the expression (\ref{eq:16}) and (\ref{eq:17}) can be reduced to the more simple form:

 \begin{equation}
 \Sigma^W(p) \simeq \sqrt 2 \, G_F \left ( \cdots - \frac{n^0_e - \tilde n^0_e}{EB} \, (p \, \tilde F \, \gamma) \right )\, \gamma_L,\label{eq:18}
 \end{equation}
 \begin{eqnarray}
 \Sigma^\Phi(p) & \simeq &  \frac{G_F}{\sqrt 2} \, \left \{ \cdots - \frac{n^0_e - \tilde n^0_e}{EB} \, (p \, \tilde F \, \gamma)\,  \right . \left ( \frac{m^2_e}{m^2_W} \, \gamma_L + \frac{m^2_\nu}{m^2_W} \, \gamma_R \,\right ) +
 \label{eq:19}\\
 & + & \left .   \frac{ie}{4\pi^2}\, m_\nu\, \frac{m^2_e}{m^2_W} \, \int\limits_{0}^{+\infty} \frac{dk}{\omega_0} \, (f(\omega_0) - \tilde f(\omega_0)) \, (\gamma \, F \, \gamma )\, \right \}, \nonumber
 \end{eqnarray}
 where ellipsis corresponds to the terms which does not contribute to the neutrino magnetic moment,  
 $\omega_0$ is the energy of electron (positron) on the ground Landau level (n=0), $n^0_e, \tilde n^0_e $  are the electron and positron densities on this level.

 Comparing the expressions  (\ref{eq:18}) and  (\ref{eq:19}) with the parametrization (\ref{eq:5}), we find the 
 coefficients $C_L, C_R$ and $K_2$:
 \begin{eqnarray}
 C_L^W & = & - \, \frac{e \, G_F}{ \sqrt 2 \, \pi^2 \, E} \, \int\limits^{+\infty}_0  dk \, (f(\omega_0) - \tilde f (\omega_0)), \label{eq:21} \\[2mm]
 C^W_R & = & K^W_2 =0, \label{eq:22}\\[2mm]
 C^\Phi_L & = & \,\frac{m^2_e}{2 m^2_W} \, C^W_L, \,\,\,\,\, C^\Phi_R = \,\frac{m^2_\nu}{2 m^2_W} \, C^W_L, \label{eq:23} \\[2mm]
 K^\Phi_2 & = & \,\frac{e \, G_F}{4 \sqrt 2 \pi^2} \, \frac{m^2_e}{m^2_W} \, \int\limits_0^{+\infty} \frac{dk}{\omega_0} \, (f(\omega_0) - \tilde f (\omega_0) ). \label{eq:24}
 \end{eqnarray}

 As one would expect, equations  (\ref{eq:23}) and (\ref{eq:24}) show that the contributions from the process due to the exchange of charges scalar $\Phi$-boson are suppressed by the factors 
 $m^2_\nu/m^2_W$ è $m^2_e/m^2_W$.
 
 As for amplitude of the neutrino forward scattering via $Z$-boson, then the lagrangian
 \begin{equation}
 L = \frac{\sqrt{g^2 + g'^2}}{4} \left ( \bar \Psi_{\nu_e} \,  \gamma_\alpha \,
( 1 - \gamma_5)\,\Psi_{\nu_e} \, \right )\,Z_\alpha + 
      \frac{\sqrt{g^2 + g'^2}}{4} \left ( \bar \Psi_{f} \,  \gamma_\alpha \,
( 1 - \gamma_5)\,\Psi_{f} \, \right )\,Z_\alpha,
       \label{eq:lag-3}
       \end{equation}
leads to the part of  $\Sigma^Z(p)$, derived from the neutrino forward scattering via $Z$-boson in the form:
 \begin{equation}
 \Sigma^Z_{f}(p)  = \sqrt 2 \, G_F \, \left ( \cdots - \frac{T^{f}_3}{BE} \, (n^0_f - \tilde n^0_f) \,
 (p\, \tilde F \,\gamma) \right )\,\gamma_L. \label{eq:26}
 \end{equation}

 Here $n^0_f$ and $ \tilde n^0_f $ are the densities of charged fermion and antifermion on ground Landau level respectively, $T_3^f$ is the third component of the isospine of a charged fermion. Ellipsis corresponds to the terms which does not contribute to the neutrino magnetic moment. Taking into account that largest density on ground Landau level corresponds to the electron and positron, from equation  (\ref{eq:26})  obtain the coefficients $C_L$, $C_R$ and $K_2$
 \begin{equation}
 C^{Z}_L =   \frac{e\,G_F}{2\sqrt 2 \, \pi^2 \,E} \, \int\limits^{+\infty}_{0} dk \, (f(\omega_0) - \tilde 
 f(\omega_0)),
 \end{equation}
 \begin{equation}
C^{Z}_R = K_2^{Z} = 0.
 \end{equation}

 So, the neutrino magnetic moment induced by the magnetized plasma is determined in the main by the coefficients 
 $C^Z_L$ and $C^W_L$ and  the field contribution
 \begin{equation}
\mu_\nu  = \mu^{plasma}_\nu + \mu^{field}_\nu \simeq \frac{m_\nu}{2}\, C_L + \mu^{field}_\nu =  \frac{m_\nu}{2} \left (C_L^W + C_L^Z \right ) + \mu_\nu^{field}. 
\end{equation}

Finally, we have 
 \begin{equation}
 \mu_\nu \simeq \frac{3eG_Fm_\nu}{8 \sqrt 2 \,\pi^2} \, 
 \left ( 1 \mp   \frac{2}{3 \,E} \, \int\limits^{+\infty}_0  dk \,
 (f(\omega_0) - \tilde f(\omega_0)) \right ), \label{eq:25}
 \end{equation}
 where the minus sign corresponds to the electron neutrino, $\nu_e$, while plus sign corresponds to the other types of neutrino, $\nu_\mu, \nu_\tau $. 
 
 The integral in expression (\ref{eq:25}) is easily calculated in  the ultrarelativistic charge asymmetric plasma
 $$ \mu_\nu = \frac{C_L\, m_\nu}{2} \simeq \frac{3e\,G_F \,m_\nu}{8 \sqrt 2 \,\pi^2} \, \left ( 1 \mp   \frac{2}{3} \, \frac{\mu}{E} \right ), $$
 and in the charge symmetric plasma
 $$     \mu_{\nu_e} \simeq \frac{3e\,G_F \,m_\nu}{8 \sqrt 2 \,\pi^2} \, \left ( 1 +  \frac{4\pi^2}{9} \, \frac{T^2}{m^2_W} \right ). $$
 
 As one can see, in contrast to \cite{Zhykovski} the neutrino magnetic moment in the magnetized plasma is suppressed by the neutrino mass. Moreover
 in the case of charge symmetric plasma neutrino magnetic moment consists additional small factor $T^2/m^2_W \ll 1$. 
 
 We believe that incorrect determination of the magnetic 
moment of the neutrino in \cite{Zhykovski} was connect with the assumption that the entire additional energy of the 
neutrino (related to its dependence on the spin and 
magnetic field) in a magnetized plasma gives a contribution to 
the induced magnetic moment. However, as it was  
shown above, only one definite part of  the 
additional neutrino energy of the neutrino refers to its magnetic 
moment.

\section{Conclusions}
\indent\indent We have investigated influence the magnetized plasma to neutrino magnetic moment under real astrophysical conditions when the mass of W-boson is the largest physical parameter.

 It is shown that
 that only a part of the additional neutrino energy depending on  its spin and magnetic field, relates to neutrino magnetic moment.  

It was found that presence of a magnetized plasma does not lead to enhancement of the neutrino magnetic moment in contrast to results in  previous literature. Moreover, the magnetic moment of the neutrino is suppressed by 
its mass, while in a charge symmetric plasma the magnetic moment it is additionally suppressed by a factor of $T/m_W \ll 1$.  

\vspace{5mm}

This work was performed in the framework of realization of the Federal
Target Program
``Scientific and Pedagogic Personnel of the Innovation Russia'' for 2009 -
2013 (State contract no. P2323)
and was supported in part by the Ministry of Education
and Science of the Russian Federation under the Program
``Development of the Scientific Potential of the Higher
Education'' (project no. 2.1.1/510).

\end{document}